\documentstyle[prd,aps,eqsecnum,epsfig]{revtex}
\newcommand{\be}{\begin{equation}}
\newcommand{\ee}{\end{equation}}

\begin{document}
\draft
\title{\bf Limits on dark matter WIMPs using upward-going 
muons in the MACRO detector}

\author{
\nobreak\bigskip\nobreak
\pretolerance=10000
\bigskip\begin{center}
{\bf The MACRO Collaboration}\\
\nobreak\bigskip\nobreak
M.~Ambrosio$^{12}$, 
R.~Antolini$^{7}$, 
C.~Aramo$^{7,n}$,
G.~Auriemma$^{14,a}$, 
A.~Baldini$^{13}$, 
G.~C.~Barbarino$^{12}$, 
B.~C.~Barish$^{4}$, 
G.~Battistoni$^{6,b}$, 
R.~Bellotti$^{1}$, 
C.~Bemporad$^{13}$, 
E.~Bernardini$^{2,7}$, 
P.~Bernardini$^{10}$, 
H.~Bilokon$^{6}$, 
V.~Bisi$^{16}$, 
C.~Bloise$^{6}$, 
C.~Bower$^{8}$, 
S.~Bussino$^{14}$, 
F.~Cafagna$^{1}$, 
M.~Calicchio$^{1}$, 
D.~Campana$^{12}$, 
M.~Carboni$^{6}$, 
M.~Castellano$^{1}$, 
S.~Cecchini$^{2,c}$, 
F.~Cei$^{11,13}$, 
V.~Chiarella$^{6}$, 
B.~C.~Choudhary$^{4}$, 
S.~Coutu$^{11,o}$,
L.~De~Benedictis$^{1}$, 
G.~De~Cataldo$^{1}$, 
H.~Dekhissi$^{2,17}$,
C.~De~Marzo$^{1}$, 
I.~De~Mitri$^{9}$, 
J.~Derkaoui$^{2,17}$,
M.~De~Vincenzi$^{14,e}$, 
A.~Di~Credico$^{7}$, 
E.~Diehl$^{11}$,
O.~Erriquez$^{1}$,  
C.~Favuzzi$^{1}$, 
C.~Forti$^{6}$, 
P.~Fusco$^{1}$, 
G.~Giacomelli$^{2}$, 
G.~Giannini$^{13,f}$, 
N.~Giglietto$^{1}$, 
M.~Giorgini$^{2}$, 
M.~Grassi$^{13}$, 
L.~Gray$^{7}$, 
A.~Grillo$^{7}$, 
F.~Guarino$^{12}$, 
P.~Guarnaccia$^{1}$, 
C.~Gustavino$^{7}$, 
A.~Habig$^{3}$, 
K.~Hanson$^{11}$, 
R.~Heinz$^{8}$, 
Y.~Huang$^{4}$, 
E.~Iarocci$^{6,g}$,
E.~Katsavounidis$^{4}$, 
I.~Katsavounidis$^{4}$, 
E.~Kearns$^{3}$, 
H.~Kim$^{4}$, 
S.~Kyriazopoulou$^{4}$, 
E.~Lamanna$^{14}$, 
C.~Lane$^{5}$, 
T.~Lari$^{2,7}$, 
D.~S. Levin$^{11}$, 
P.~Lipari$^{14}$, 
N.~P.~Longley$^{4,l}$, 
M.~J.~Longo$^{11}$, 
F.~Maaroufi$^{2,17}$,
G.~Mancarella$^{10}$, 
G.~Mandrioli$^{2}$, 
S.~Manzoor$^{2,m}$, 
A.~Margiotta Neri$^{2}$, 
A.~Marini$^{6}$, 
D.~Martello$^{10}$, 
A.~Marzari-Chiesa$^{16}$, 
M.~N.~Mazziotta$^{1}$, 
C.~Mazzotta$^{10}$, 
D.~G.~Michael$^{4}$, 
S.~Mikheyev$^{4,7,h}$, 
L.~Miller$^{8}$, 
P.~Monacelli$^{9}$, 
T.~Montaruli$^{1}$ \cite{cora}, 
M.~Monteno$^{16}$, 
S.~Mufson$^{8}$, 
J.~Musser$^{8}$, 
D.~Nicol\'o$^{13,d}$,
C.~Orth$^{3}$, 
G.~Osteria$^{12}$, 
M.~Ouchrif$^{2,17}$,
O.~Palamara$^{7}$, 
V.~Patera$^{6,g}$, 
L.~Patrizii$^{2}$, 
R.~Pazzi$^{13}$, 
C.~W.~Peck$^{4}$, 
S.~Petrera$^{9}$, 
P.~Pistilli$^{14,e}$, 
V.~Popa$^{2,i}$, 
A.~Rain\`o$^{1}$, 
A.~Rastelli$^{2,7}$, 
J.~Reynoldson$^{7}$, 
F.~Ronga$^{6}$, 
A.~Sanzgiri$^{15,p}$,
C.~Satriano$^{14,a}$, 
L.~Satta$^{6,g}$, 
E.~Scapparone$^{7}$, 
K.~Scholberg$^{3}$, 
A.~Sciubba$^{6,g}$, 
P.~Serra-Lugaresi$^{2}$, 
M.~Severi$^{14}$, 
M.~Sioli$^{2}$, 
M.~Sitta$^{16}$, 
P.~Spinelli$^{1}$, 
M.~Spinetti$^{6}$, 
M.~Spurio$^{2}$, 
R.~Steinberg$^{5}$,  
J.~L.~Stone$^{3}$, 
L.~R.~Sulak$^{3}$, 
A.~Surdo$^{10}$, 
G.~Tarl\`e$^{11}$,   
V.~Togo$^{2}$, 
D.~Ugolotti$^{2}$, 
M.~Vakili$^{15}$, 
C.~W.~Walter$^{3}$,  and R.~Webb$^{15}$.\\
\vspace{1.5 cm}
\footnotesize
1. Dipartimento di Fisica dell'Universit\`a di Bari and INFN, 70126 
Bari,  Italy \\
2. Dipartimento di Fisica dell'Universit\`a di Bologna and INFN, 
 40126 Bologna, Italy \\
3. Physics Department, Boston University, Boston, MA 02215, 
USA \\
4. California Institute of Technology, Pasadena, CA 91125, 
USA \\
5. Department of Physics, Drexel University, Philadelphia, 
PA 19104, USA \\
6. Laboratori Nazionali di Frascati dell'INFN, 00044 Frascati (Roma), 
Italy \\
7. Laboratori Nazionali del Gran Sasso dell'INFN, 67010 Assergi 
(L'Aquila),  Italy \\
8. Depts. of Physics and of Astronomy, Indiana University, 
Bloomington, IN 47405, USA \\
9. Dipartimento di Fisica dell'Universit\`a dell'Aquila  and INFN, 
 67100 L'Aquila,  Italy \\
10. Dipartimento di Fisica dell'Universit\`a di Lecce and INFN, 
 73100 Lecce,  Italy \\
11. Department of Physics, University of Michigan, Ann Arbor, 
MI 48109, USA \\	
12. Dipartimento di Fisica dell'Universit\`a di Napoli and INFN, 
 80125 Napoli,  Italy \\	
13. Dipartimento di Fisica dell'Universit\`a di Pisa and INFN, 
56010 Pisa,  Italy \\	
14. Dipartimento di Fisica dell'Universit\`a di Roma ``La Sapienza" and INFN, 
 00185 Roma,   Italy \\ 	
15. Physics Department, Texas A\&M University, College Station, 
TX 77843, USA \\	
16. Dipartimento di Fisica Sperimentale dell'Universit\`a di Torino and INFN,
 10125 Torino,  Italy \\	
17. L.P.T.P., Faculty of Sciences, University Mohamed I, 
B.P. 524 Oujda, Morocco \\
$a$ Also Universit\`a della Basilicata, 85100 Potenza,  Italy \\
$b$ Also INFN Milano, 20133 Milano, Italy\\
$c$ Also Istituto TESRE/CNR, 40129 Bologna, Italy \\
$d$ Also Scuola Normale Superiore di Pisa, 56010 Pisa, Italy\\
$e$ Also Dipartimento di Fisica, Universit\`a di Roma Tre, Roma, Italy \\
$f$ Also Universit\`a di Trieste and INFN, 34100 Trieste, 
Italy \\
$g$ Also Dipartimento di Energetica, Universit\`a di Roma, 
 00185 Roma,  Italy \\
$h$ Also Institute for Nuclear Research, Russian Academy
of Science, 117312 Moscow, Russia \\
$i$ Also Institute for Space Sciences, 76900 Bucharest, Romania \\
$l$ Swarthmore College, Swarthmore, PA 19081, USA\\
$m$ RPD, PINSTECH, P.O. Nilore, Islamabad, Pakistan \\
$n$ Also INFN Catania, 95129 Catania, Italy\\
$o$ Also Department of Physics, Pennsylvania State University, 
University Park, PA 16801, USA\\
$p$ Tektronix Inc., Wilsonville, OR\\
\end{center}
}
\date{\today}
\maketitle
\vskip 0.5 cm
Submitted to Phys. Rev. D
\vskip 0.5 cm

\begin{abstract}
We perform an indirect search for Weakly Interacting Massive Particles
(WIMPs) using the MACRO detector to look for neutrino-induced upward-going
muons resulting from the annihilation of WIMPs trapped in the Sun and
Earth. The search is conducted in various angular cones centered
on the Sun and Earth to accommodate a range of WIMP masses.  No
significant excess over the background from atmospheric neutrinos is
seen and limits are placed on the upward-going muon fluxes from Sun and
Earth. These limits are used to constrain neutralino particle
parameters from supersymmetric theory, including those suggested by recent
results from DAMA/NaI.
\end{abstract}


\section{Introduction}

There are many hints for the existence of non-baryonic dark matter in
our universe \cite{Primack97}, which may consist of {\it Weakly
Interacting Massive Particles} (WIMPs).  
Measurements of the mass density of the universe, $\Omega_{M}$, 
indicate a density considerably in excess of the density of baryonic matter 
allowed by Big Bang nucleosynthesis (BBN) calculations.  
For example, the POTENT analysis of peculiar velocities of galaxies 
\cite{Dekel94} excludes $\Omega_{M} < 0.3$ at the 2.4$\sigma$ level 
($\Omega_{M}$ is in units of the critical density).  
In contrast, the limits for baryonic mass ($\Omega_{b}$)
from BBN are $0.005 \lesssim \Omega_{b} \lesssim 0.10$ at 95$\%$ c.l.
\cite{Olive96} for $0.4 \lesssim h \lesssim 1$, 
where $h$ is the scaled value of the Hubble constant.  
This discrepancy is an indication
that there must be some undiscovered non-baryonic dark matter.
Large-scale structure models coupled with COBE data \cite{COBE} also
favor cosmological scenarios with large amounts of dark matter. One
recent model favors a mixture of 70$\%$ Cold Dark Matter, 20$\%$ Hot Dark
Matter and 10$\%$ baryons \cite{Primack97}. At smaller scales (roughly
100 to 10000 kpc) virial estimates on groups and clusters of galaxies
and rotation curves of spiral galaxies \cite{Salucci} are consistent
with smaller values of $\Omega_{M}$. Furthermore, dark matter in 
galaxies is also motivated by the fact that dark matter halos seem to 
help stabilize spiral disk structure \cite{Ostriker73}.
Nevertheless, the amount of matter indicated cannot be accounted for 
by baryonic matter: MACHOs are unable to account for all of the 
dark matter halo of our galaxy \cite{Alcock97}.

A long list of Cold Dark Matter nonbaryonic candidates has been
suggested, among which the supersymmetric (SUSY) neutralino,
considered in this paper, and the axion seem to be the most promising
\cite{Jungman96}. SUSY postulates a symmetry between bosons and
fermions predicting SUSY partners to all known particles (for a review
see Ref.~\cite{Kane85}). SUSY solves a host of particle physics
questions such as the ``hierarchy problem'' (explaining the large
difference between the weak and GUT scales); generating electroweak
symmetry breaking through the Higgs mechanism; and stabilizing the
Higgs mass at the weak scale. In theories where R-parity is conserved
there exists a stable Lightest Supersymmetric Particle (LSP). If the
neutralino is the LSP it is a natural WIMP candidate: it is a weakly
interacting particle with a mass between roughly a GeV and a TeV and
would be expected to have a significant relic density.

The neutralino $\tilde \chi$ is the lightest linear
superposition of gaugino and higgsino eigenstates
(gauginos and higgsinos are SUSY counterparts to the Higgs and gauge bosons):

\begin{equation}
\tilde \chi  \; \; = \; \; a_{1}\tilde{\gamma} + a_{2}\tilde{Z} +
a_{3}\tilde{H_{1}} + a_{4}\tilde{H_{2}}  
\end{equation}

\noindent where $\tilde{\gamma}$ and $\tilde{Z}$ are gaugino states;
$\tilde{H_{1}}$ and $\tilde{H_{2}}$ are higgsino states.  
If the minimal supersymmetric extension of the standard model (MSSM)
\cite{Jungman96} is coupled with some GUT assumptions ($M_{1} = 5/3
M_{2} \sin^2 \theta_{W}$ where $\theta_{W}$ is the Weinberg angle),
then the neutralino mass depends on one of the two gaugino mass
parameters, $M_{1}$ or $M_{2}$, on the higgsino mass parameter $\mu$,
and on the ratio of the Higgs doublet vacuum expectation values, $\tan
\beta$. The phenomenology of neutralinos is determined by the
composition parameter $P = a_{1}^{2} + a_{2}^{2}$. Some other
parameters must be determined in order to define the processes induced
by neutralinos, because Higgs and the supersymmetric partners of
fermions ({\it sfermions}) play a relevant role. In the MSSM there are
two Higgs doublets, hence three neutral Higgs fields (two scalar and
one pseudoscalar). The Higgs sector is determined by two independent
parameters: $\tan\beta$ and $m_{A}$, the mass of the pseudoscalar
neutral boson. The other parameters in the Lagrangian of the model
are the bilinear and trilinear parameters connected to the spontaneous
symmetry breaking. The number of parameters needed to describe
neutralino phenomenology may be reduced further by assuming that all
the trilinear parameters are zero except for the third family, which
are assumed to have the common value, $A$, and that all the squarks and
sleptons are degenerate with common mass, $m_{0}$. In this paper we
use a model developed following the assumptions above \cite{Bottino98}.

Currently, supersymmetric parameter space is constrained by
accelerator searches.  Data from LEP \cite{LEP2}
lead to lower limits on the neutralino mass, $m_{\chi}$, between
20-30 GeV. These limits are model-dependent and correlated to limits
from chargino (a mixture of w-inos and charged higgsinos) searches.

In this framework, ``direct'' and ``indirect'' methods for detecting
Galactic halo WIMPs can probe complementary regions of the
supersymmetric parameter space, even when more extensive LEP 2 results
become available.  Direct methods detect WIMPs via a direct
interaction of a WIMP, such as by observing the energy deposited in a
low-background detector (e.g., semiconductors or scintillators) when a
WIMP elastically scatters from a nucleus.  Indirect methods look for
by-products of WIMP decay or annihilation such as neutrinos resulting
from the annihilation of WIMPs.  An excellent prospect for indirect
WIMP searches is to look for high energy neutrinos from WIMP
annihilation in the core of the Earth or the Sun
\cite{Freese86,Krauss86,Silk85}. In MACRO, such neutrinos would be
detected as neutrino-induced upward-going muons which can be
distinguished from downward-going cosmic-ray shower muons.

\section{Indirect search through upward-going muons}

Dark matter WIMPs in the Galactic halo can be captured in a celestial
body by losing energy through elastic collisions and becoming
gravitationally trapped. As the WIMP density increases in the core of
the body, the WIMP annihilation rate increases until equilibrium is
achieved between capture and annihilation. High energy neutrinos are
produced via the hadronization and decay of the annihilation products
(mostly fermion-antifermion pairs, weak and Higgs bosons) and may be
detected as upward-going muons in underground detectors.

There are many calculations of the expected neutrino fluxes from WIMP
capture and annihilation in the Sun and Earth
\cite{Jungman96,Bottino95,Bergstrom97}.  
The capture rate for an astrophysical body depends on several factors: 
the WIMP mean halo velocity ($\sim 270$ km s$^{-1}$); the WIMP local density
($\rho_{\chi} \sim 0.3-0.6$ GeV cm$^{-3}$); the WIMP scattering
cross-section; and the mass and escape velocity of the celestial body
\cite{Jungman96}. The WIMP may scatter from nuclei with spin
(e.g. hydrogen in the Sun) via an axial-vector (``spin-dependent'')
interaction in which the WIMP couples to the spin of the nucleus or
via a scalar interaction in which the WIMP couples to the nuclear
mass. In axial-vector interactions the probability for a given energy
loss is constant up to the kinematic limit of the interaction, while
for a scalar interaction there will be a suppression of the cross
section at high momentum transfers \cite{Jungman96}. 
Elastic scattering is most
efficient when the mass of the WIMP is similar to the mass of the
scattered nucleus. Hence, the heavy nuclei in the Earth make it very
efficient in capturing WIMPs with $m_{\chi} \lesssim 100$ GeV (the
resonance effect \cite{Gould}). The Sun, in contrast, has a smaller
average nuclear mass, but is nonetheless efficient in capturing WIMPs
due to its larger escape velocity.
 
A WIMP annihilation signal would appear as a statistically significant
excess of upward-going muon events from the direction of the Sun or of the
Earth among the background of atmospheric neutrino-induced upward-going
muons. High energy neutrino-induced upward-going muons tend to retain
the directionality of the parent neutrino. This directionality
permits a restriction of the search for WIMP annihilation neutrinos to
a narrow cone pointing from the Earth or Sun, greatly reducing the
background from atmospheric neutrinos. This effect, along with the
increase in neutrino cross-section with energy and longer range of
high energy muons, means that this method of detection achieves an
increasingly better signal to noise ratio for high WIMP masses.
 
Data on upward-going muons from the core of the Earth and of the Sun
have been presented by several experiments, notably Baksan
\cite{baksan96}, Kamiokande \cite{Mori93} and IMB \cite{Losecco87}.
In this paper we describe a WIMP search using the MACRO detector.

\section{WIMP search in MACRO from the Earth and the Sun}

The MACRO apparatus \cite{Ahlen93}, located in the Gran Sasso
Underground Laboratory of the Italian {\it Istituto Nazionale di Fisica 
Nucleare},
detects upward-going muons using a system of limited streamer tubes for
tracking (angular resolution $\sim 0.5^{o}$) and roughly 600 tons of
liquid scintillator for fast timing (time resolution $\sim 500$ ps).
The detector has overall dimensions of $\rm 12 \times 77 \times 9
\;m^{3}$ and it is divided in 6 parts called {\it supermodules}. 
The bottom part of the apparatus, 4.8 m high, is filled with
rock absorber which sets a minimum threshold of about 1 GeV for
vertical muons crossing the detector. The upper part, called the {\it
attico}, is an open volume containing electronics as well as active
detector elements.  The streamer tubes form 14 horizontal and 12
vertical planes and the liquid scintillator counters form 3 horizontal
and 4 vertical planes on the outer surfaces of the detector.

Neutrino events in the MACRO detector are seen 
in three different topologies:

\begin{enumerate}
\item through-going upward muons. These events are produced by 
neutrinos interacting in the rock below MACRO and pass entirely
through the detector. For atmospheric neutrinos the 
spectrum ranges from 1 to 10$^{4}$ GeV and the peak 
energy is about 100~GeV; 
\item internally produced upward-going events induced by neutrinos 
interacting in the lower part of the detector and producing a lepton 
which moves upward through the two upper scintillator layers. 
For atmospheric neutrinos the peak energy is about 4 GeV.
To detect these events the {\it attico} must be in operation;
\item externally produced neutrino-induced upward-going stopping muons 
and internally-produced downward-going leptons.  
For atmospheric neutrinos the peak energy is about 4 GeV.
\end{enumerate}

The first and second topologies are recognized using the time-of-flight 
technique. Topological criteria are used for the third category.
The time-of-flight technique is used to discriminate upward-going
neutrino-induced muons from the background of downward-going
atmospheric muons. Each scintillator records the time of a particle
crossing by measuring the mean time at which signals are observed
at the two ends of the scintillator. The transit time of a particle
is found by taking the difference between the crossing times of two
scintillators. In our convention, upward-going muons are considered to have
a negative velocity, and downward-going muons a positive velocity.

We present the results of WIMP searches using data gathered from March
1989 to March 1998. This data set encompasses five years of running with 
partial apparatus (March 1989 to April 1994) and 4 years of running
with the full apparatus including the attico (April 1994 to March
1998) and it corresponds to livetimes including efficiency
of 1.38 yr of running with the $1^{st}$ {\it supermodule}, 0.41 yr 
of running with the lower part of the detector and 3.1 yr 
of running of the full detector.
The details of the upward-going muon analysis for the $1^{st}$ 
supermodule and the lower detector can be found in
Ref.~\cite{MACRO95}, for the full apparatus in Ref.~\cite{Letter}.
    
Slightly different data sets and selection criteria are used for WIMP
searches for the Earth and Sun to optimize the signal over background
for each. For the Earth analysis only through-going upward muons
which traverse at least 200 g/cm$^2$ of absorber are used (category 1
above). The absorber requirement reduces the background due to soft
pions produced at large angles by undetected downward-going muons to
1$\%$ \cite{Spurio}. A total of 517 through-going upward muon events
are used in the search for WIMP annihilation neutrinos from the Earth.
    
Background rejection is not so critical for moving sources as it is
for steady sources, so the analysis for the Sun includes through-going
upward muons with no absorber requirement as well as 
semi-contained events generated in the lower half of the detector (category 2 
above), for a total of 762 upward-going muon events. Apart from this,
the analysis for the Sun is basically the same as the one for the Earth.

We consider several contributions to the background of
through-going upward muons: events with an incorrect timing
measurement (such as muons in coincidence with radioactivity, other
muons, or electromagnetic showers) and soft pions produced at large
angles. We estimate 20 background events in the Earth sample of 517 
through-going upward muons.

The expected background of upward-going muons from atmospheric neutrinos is
calculated with a full Monte Carlo calculation (described in
Ref.~\cite{MACRO95,Letter}). This calculation uses the Bartol flux
\cite{Agrawal}, the Morfin and Tung parton set S$_{1}$ \cite{Morfin}
for the deep inelastic $\nu N$ cross-section and the muon energy loss
in the rock from Ref.~\cite{Lohmann}. We estimate a total uncertainty
in the calculation of 17$\%$ \cite{MACRO95,Letter}. We have considered
scenarios both with no neutrino oscillations as well as with the
neutrino oscillation parameters of Ref.~\cite{Letter}. The
oscillations parameters considered are $\Delta m^{2} = 2.5 \times
10^{-3}$ eV$^{2}$ and $\sin^{2}2\theta = 1$. 

In the no oscillation scenario the expected number of atmospheric
neutrino events is $662 \pm 113_{theor}$  
as compared to $462 \pm 79_{theor}$ in the oscillation scenario.  
In both cases there is a deficit of the measured events with respect to 
the expected number in the region around the vertical direction of MACRO, 
where the efficiency and acceptance of the apparatus are best known. 
Because of the discrepancy between the observed and expected numbers of 
upward-going muons 
in the expected signal region ($0^\circ$ to $30^\circ$ from the
vertical), we normalize the expected signal using data outside the
expected signal region. This normalization is motivated by the fact
that the absolute error of the expected flux is relatively high,
whereas the shape of the flux is known to a few percent
\cite{Lipari98}. This normalization factor is determined separately
for each of the search cones considered, using the ratio of observed
to expected events outside the search cone.

\begin{center}
\begin{figure}[htb]
\epsfig{file=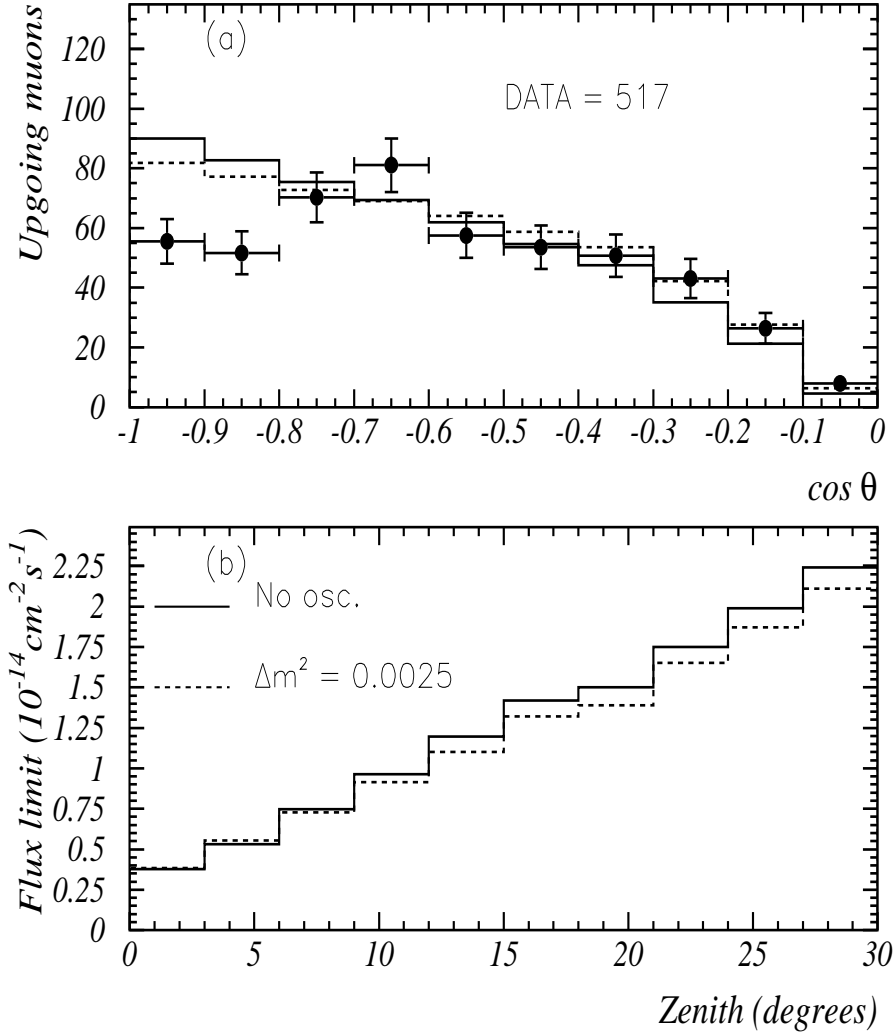,height=16.cm,width=13.cm}
\caption{(a) Nadir distribution of measured (black circles)
and expected (solid line and dashed line) through-going upward muons.
The expected distributions are multiplied by the ratio of the measured
events over the expected ones outside the largest window ($30^{\circ}$).
The normalization factors are 0.82 for the solid line (no oscillations)
and 1.19 for the dashed line ($\nu_{\mu}-\nu_{\tau}$ oscillations for
maximum mixing and $\Delta m^{2} = 0.0025$ eV$^{2}$).
(b) Muon flux limits (90$\%$ c.l.) as a function of the angle from the 
vertical (the angle varies form $3^{\circ}$ to 30$^{\circ}$ in steps
of $3^{\circ}$). 
In both plots, the dashed line is obtained in the hypothesis
of $\nu_{\mu} \rightarrow \nu_{\tau}$ oscillations of the atmospheric neutrino
background with maximum mixing and $\Delta m^{2} = 0.0025$ eV$^{2}$.}
\label{fig1}
\end{figure}
\end{center}

We show both the expected atmospheric neutrino background with and
without oscillations in Fig.~\ref{fig1}(a) and (b).
If neutrinos oscillate with these parameters the expected number of
events would be reduced and the angular distribution of through-going
upward muons would be distorted because neutrinos at the nadir
oscillate more than at the horizon due to the longer pathlength
(for a full discussion see Ref.~\cite{Letter}). 
We note that the slight excess observed in the region $-0.7 \lesssim
\cos\theta \lesssim -0.6$ is very unlikely to be produced by any
plausible WIMP model, since this region is far from the expected
signal region ($\theta \lesssim 30^\circ$ - see Fig.~\ref{fig3}(a)).

Muon flux limits are evaluated as:

\begin{equation}
\rm \Phi(90\% \, c.l.) = \frac{N_{\cal P}(90\% \, c.l.)}{\cal E}
\end{equation}

\noindent where $\rm N_{\cal P}$ is the upper Poissonian limit (90$\%$
c.l.) given the number of measured events and expected background
\cite{PDB} due to atmospheric neutrinos 
and $\cal E$ is the exposure given by equation:

\begin{equation}
\label{EXP}
\rm {\cal E} = \int_{T_{start}}^{T_{end}} {\epsilon (t) \times A(\Omega(t)) dt}
\end{equation}

\noindent where A($\Omega$(t)) is the detector area in the direction
of the expected signal ($\Omega$) at time t; $\epsilon$ is the
detector efficiency (discussed in Ref.~\cite{MACRO95,Letter})
which takes into account the possible variations of detector running 
configuration during data taking; and
$\rm T_{start}$ and $\rm T_{end}$ are the start and end times of data
taking. 
In the case of the Earth, the signal is expected always from the same
direction; hence $\Omega$ is a constant with time and the detector
acceptance $A$ is calculated using the Monte Carlo calculation in 
Ref.~\cite{MACRO95,Letter}; the live-time and efficiency are calculated 
using downward-going events.

Since the number of detected events is less than expected from
the atmospheric neutrino flux, we set conservative flux limits assuming that 
the number of measured events in the signal region equals the number of 
expected events in that region \cite{PDB}.
In Fig.~\ref{fig1}(b) the 90$\%$ c.l. muon flux limits for the Earth 
are plotted as a function of the nadir angle for 10 search half-cones 
around the vertical from $3^{\circ}$ to $30^{\circ}$.
For this data, the average value of the exposure is 2620 m$^{2}$ yr (it
varies slightly for different search cones because the area decreases
by 37 m$^{2}$ when the search cone increases from 3$^{\circ}$ to
30$^{\circ}$ around the vertical).
Note that the flux limits are independent on any hypothesis on a WIMP
signal (or any other source). As shown in Fig.~\ref{fig1}(b), 
the application of a $\nu_{\mu}-\nu_{\tau}$
oscillation hypothesis to the atmospheric neutrino background
with $\Delta m^{2} = 0.0025$ eV$^{2}$ 
and maximum mixing will result in lower flux limits. 

\begin{center}
\begin{figure}[htb]
\epsfig{file=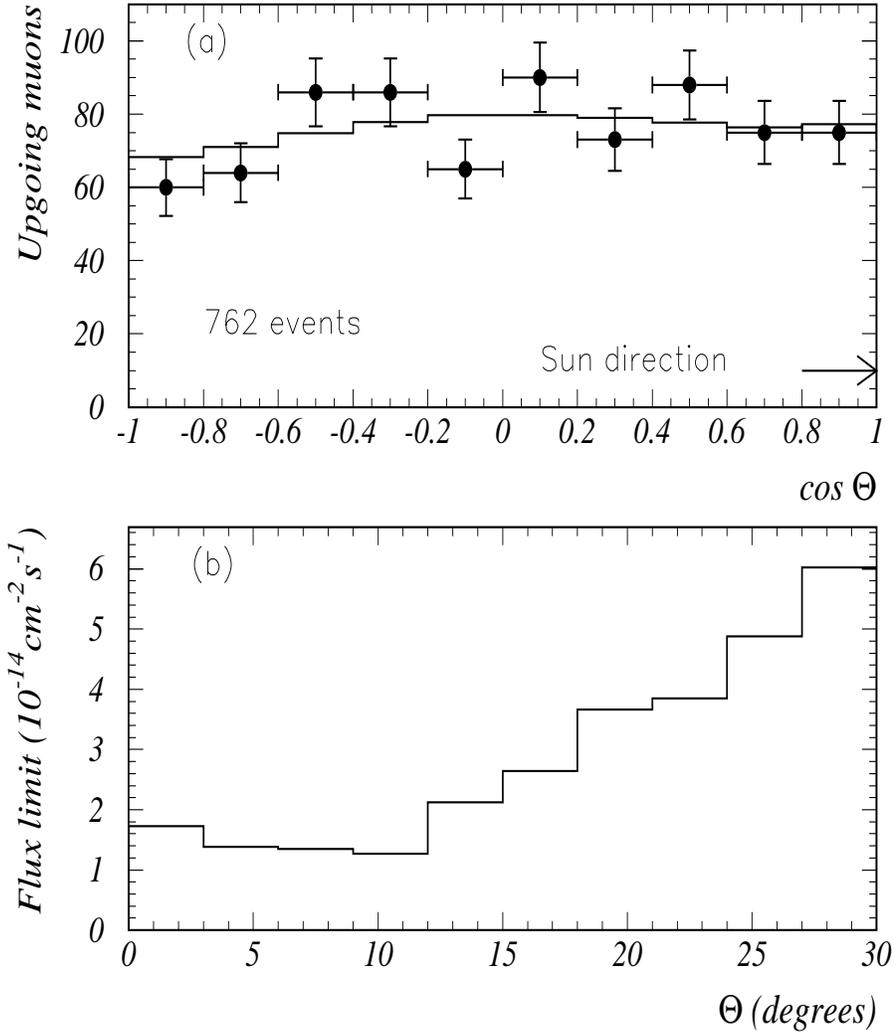,height=16.cm,width=13.cm}
\caption{(a) Distribution of measured (black circles) and expected
(solid line) upward-going muons as a function of the cosine of the
angle from the Sun direction.  (b) Muon flux limits (90\% c.l.) as a
function of the search cone around the direction of the Sun.}
\label{fig2}
\end{figure}
\end{center}

\begin{center}
\begin{figure}[htb]
\epsfig{file=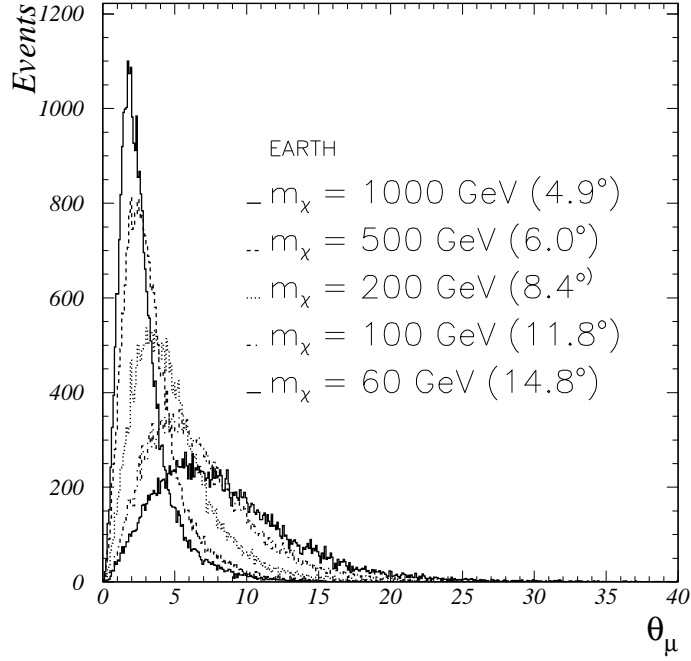,height=10.cm,width=10.cm}
\caption{Nadir angle distribution of muons induced by neutralino
annihilation inside the Earth for several neutralino masses.
The angular ranges including 90$\%$ of the signal are indicated.}
\label{fig3}
\end{figure}
\end{center}

In the analysis for muons pointing in the direction of the Sun, 
the data themselves are used to generate
the expected events in order to properly include the effects of
semi-contained events. The arrival times from measured downward-going
muons from the entire period of data taking are assigned randomly to
the local trajectory coordinates of the measured upward-going muon events to
evaluate their right ascension. This procedure allows us to take into
account drifts of detection efficiency in time. Fig.~\ref{fig2}(a)
shows the angular distributions of measured and expected upward-going muons
with respect to the direction of the Sun.  The shape depends on the
seasonal variation of the position of the Sun and on the livetime of
the apparatus. The upward-going events detected during the night fall
towards a cosine of 1, while the events collected during the day fall
near -1. The exposure of the Sun is calculated with
eq.~\ref{EXP} using the detector acceptance from the Monte Carlo
described in Ref.~\cite{Letter} (for $E_{\mu} > 2$ GeV, see below).
The dependence of the acceptance of the detector on the direction of the 
Sun when it is below the horizon is exactly calculated using both 
the Monte Carlo and downward-going muon data.

In Fig.~\ref{fig2}(b) the flux limits (90$\%$ c.l.) for an exposure of
$\sim$ 890 m$^{2}$ yr (it varies slightly for different 
search cones) for 10 search cones around the Sun direction are shown.  
Again, it should be noted that any neutrino oscillation effects 
are automatically 
included in our method of deriving the number of expected events from real
data.

Table~\ref{tab1} shows the number of detected and expected
upward-going muon events and corresponding flux limits for several
cone sizes for the Earth and Sun. The muon threshold energy is
determined by the amount of absorber an upward-going muon must
traverse in MACRO. It is lower for the Earth (where tracks are
oriented towards the vertical) than it is for the Sun (where tracks
are more inclined). 
The upper limits in Tab.~\ref{tab1} are calculated assuming a minimum
energy of 1.5 GeV and 2 GeV for the Earth and the Sun, respectively.
Given the analysis requirements to select upward-going muons
above these energies, we make a maximum error of 5$\%$ with respect
to an exact calculation which takes into account the dependence on energy 
of the acceptance of the apparatus and of the
neutrino fluxes from neutralinos with different masses. 


\section{Flux limits compared to SUSY predictions}

The flux limits from the previous section can be used to constrain any
WIMP model. In this section, we apply our flux limits to the
neutralino models of Bottino {\it et al.} \cite{Bottino98} using
search cones which collect 90$\%$ of the expected signal.  The range of
search cones accommodates a range of neutralino masses.

The angular distribution of upward-going muons follows that of the parent
neutrinos,
with deviations due to charged-current neutrino interaction and Coulomb
multiple scattering of the produced muons in their path to the detector.
Both of these distributions are a function of the neutrino energy. In
turn, the energy spectrum of neutrinos produced in WIMP annihilation
depends both on the WIMP mass and the annihilation products. The final
states can be pairs of fermions, or Higgs/gauge bosons or combinations
of Higgs and gauge bosons. The branching ratios into these channels
depend on the model and they have some effect on the neutrino energy
spectra. Annihilation to fermion pairs tends to produce softer
neutrinos than annihilation to gauge bosons since fermions dissipate
energy in hadronization whereas bosons have a greater likelihood of
prompt decay to neutrinos. We estimate the maximum variation in flux
limits by taking the extreme cases where only one annihilation channel
is open. We find this variation is no more than 17$\%$ between
extreme models and those considered here. To a good approximation,
the angular distribution is a function of just the neutralino mass,
and so we determine flux limits as a function of neutralino mass.

The angular distributions of the upward-going muon 
signals are calculated using
neutrino fluxes from neutralino annihilation in the Sun and Earth
calculated by Bottino {\it et al.}~\cite{Bottino98}. A Monte Carlo
calculation with cross-sections described in Ref.~\cite{Lipari95} and
muon energy loss as described in Ref.~\cite{Lipari91} is used to
propagate muons through the detector where the angular resolution is
taken into account.

The main difference between the
angular distributions of the signals from Earth and
Sun is due to the angular resolution of the detector which degrades
for slanted tracks because the number of streamer tube layers crossed
decreases. Hence, the more slanted tracks from the Sun tend to have
lower angular resolution than those from the Earth.
Moreover, the angular size of the expected signal is affected by the angular
size of the region where WIMP annihilation is taking place. The
diameter of the Sun is $0.5^{\circ}$ (as seen by the Earth) and the
WIMPs are expected to be localized at the center of the Sun. As a
result, the angular size of the WIMP annihilation region is
negligible. However, for the Earth the angular size of the
annihilation region is considerable and has been estimated by
\cite{Griest,Gould,Bottino95} to be:

\begin{equation}
G(\theta) \simeq 4 m_{\chi} \alpha e^{-2 m_{\chi} \alpha \sin^{2}\theta}
\end{equation}

\noindent where $\theta$ is the nadir angle; $\alpha$ is a
parameter depending on the central temperature (T = 6000 K), the
central density ($\rho = 13$ g cm$^{-3}$) and radius of the Earth
($\alpha = 1.76$ GeV$^{-1}$).

\begin{center}
\begin{figure}[htb]
\epsfig{file=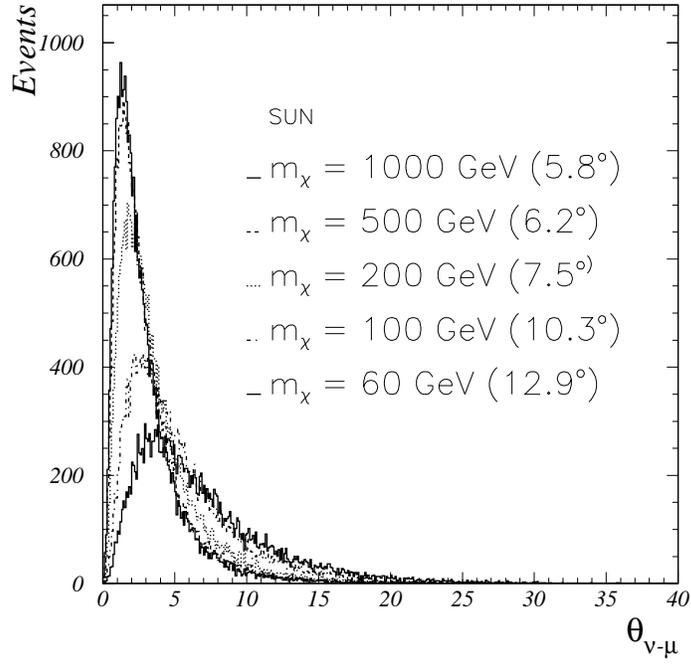,height=10.cm,width=10.cm}
\caption{Neutrino-muon angular separation distribution for neutrinos
from $\tilde \chi - \tilde \chi$ annihilation in the Sun for several neutralino
masses. The angular ranges with 90$\%$ of the signal are shown.}
\label{fig4}
\end{figure}
\end{center}

In Fig.~\ref{fig3} the muon nadir angle is shown for $m_{\chi}$ = 60,
100, 200, 500 and 1000 GeV and fixed model parameters. In
Fig.~\ref{fig4} the angular spreads between the neutrino
and the muon directions are shown as a function of neutralino
mass in the case of the search for neutralinos trapped inside the Sun.

\begin{center}
\begin{figure}[htb]
\epsfig{file=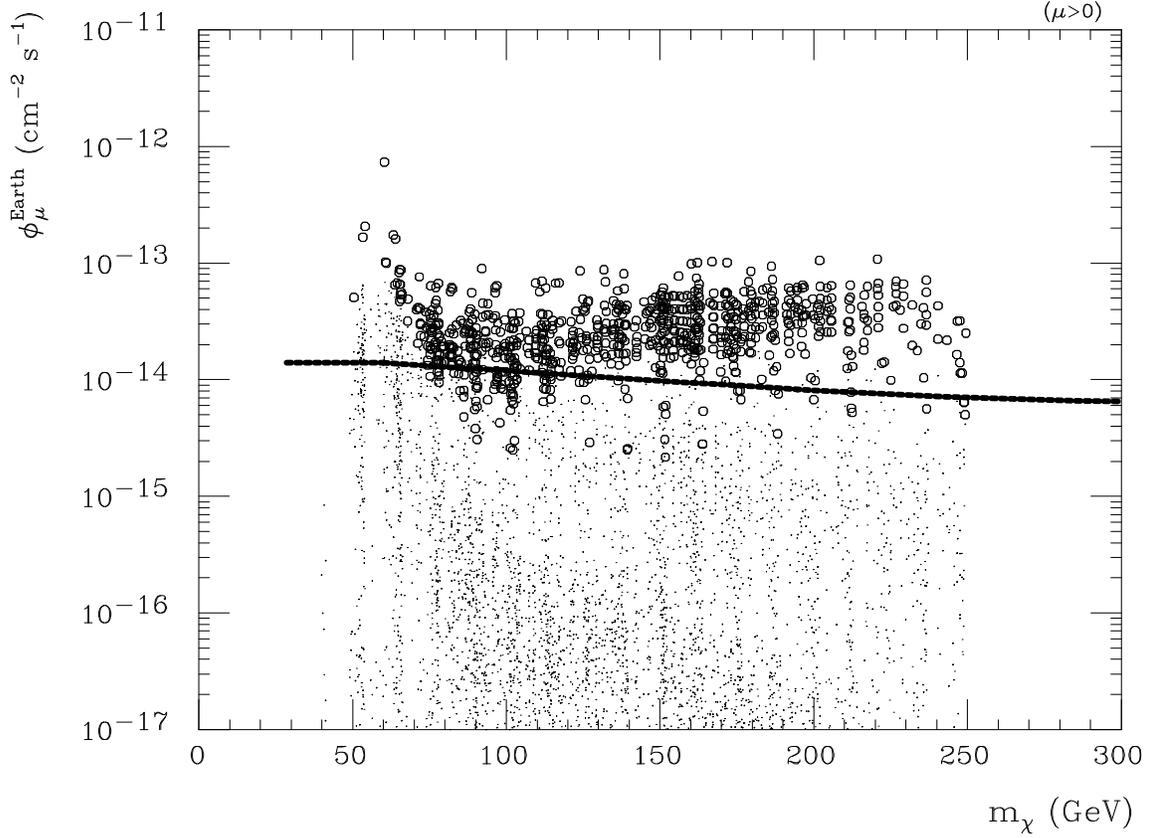,height=16.cm,width=13.cm}
\vskip -.05cm
\caption{Upward-going muon flux vs $m_{\chi}$ for $E_{\mu}^{\rm th} =$ 1
GeV from the Earth \protect\cite{Bottino98}. Each dot is obtained
varying model parameters. In this plot values of $\mu > 0$ are considered. 
Similar results are obtained for $\mu <0$.
Solid line: MACRO flux limit (90$\%$ c.l.).
The solid line representing the flux limit for the no-oscillation
hypothesis is indistinguishable in the log scale from the one for the 
$\nu_{\mu}-\nu_{\tau}$ oscillation hypothesis, but the expectations
could be about two times lower.
The open circles indicate the models {\it excluded} by
direct measurements (particularly the DAMA/NaI experiment
\protect\cite{DAMA96}) and assume a local dark matter density of 0.5
GeV cm$^{-3}$. See Fig.~\ref{fig7} for the comparison for the same density
between the MACRO flux limit and the allowed values of parameters based on 
recent DAMA/NaI results.}
\label{fig5}
\end{figure}
\end{center}

\begin{center}
\begin{figure}[htb]
\epsfig{file=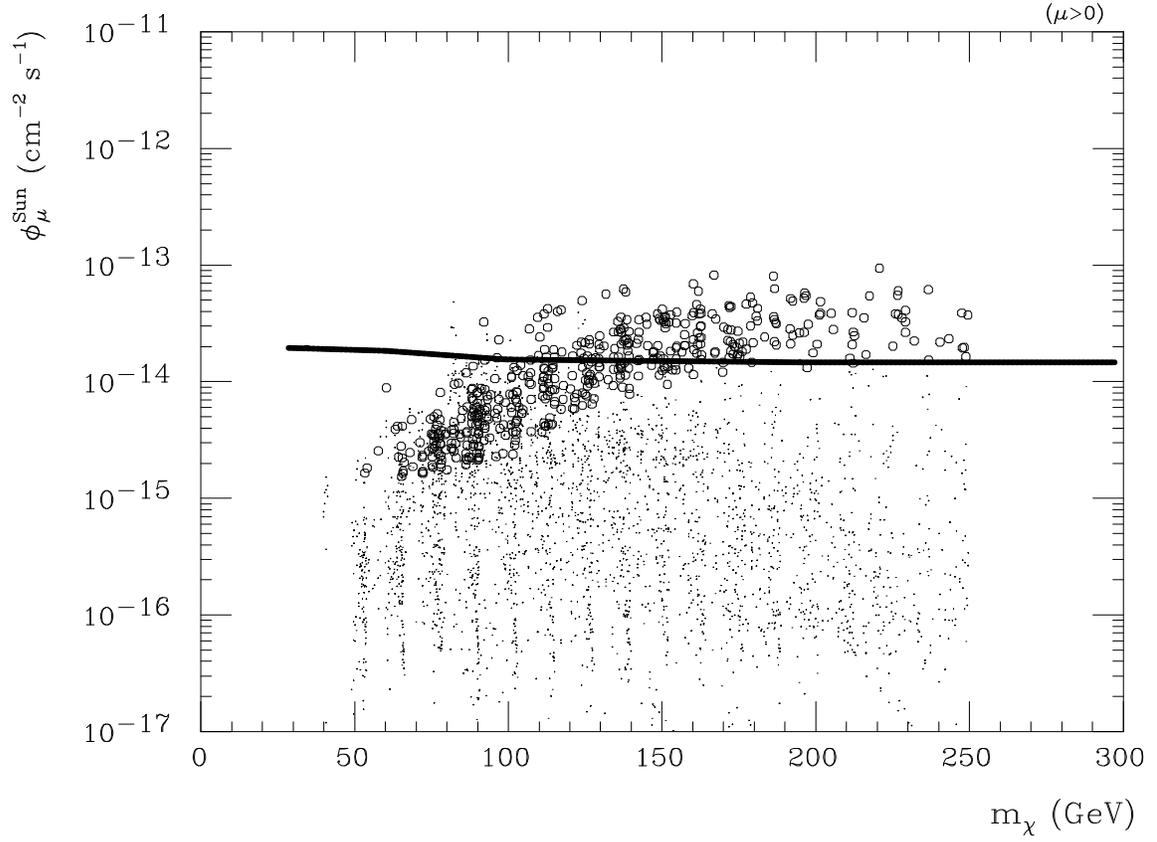,height=16.cm,width=13.cm}
\vskip -.05cm
\caption{Upward-going muon flux vs $m_{\chi}$ for $E_{\mu}^{\rm th}$ = 1 GeV
from the Sun \protect\cite{Bottino98}. In this plot values of $\mu > 0$ are 
considered. 
Solid line: MACRO flux limit (90$\%$ c.l.).
The open circles concern the regions excluded by direct measurements
\protect\cite{DAMA96}.}
\label{fig6}
\end{figure}
\end{center}

The 90$\%$ c.l. flux limits are calculated as a function of neutralino
mass using cones which collect 90$\%$ of the expected signal. These
limits are corrected for the 90$\%$ collection efficiency due to cone
size. Figures~\ref{fig5} and \ref{fig6} show these limits for the
Earth and Sun, respectively. The experimental limits are superimposed
on the flux of upward-going muons from the Bottino {\it et al.}
calculation as a function of $m_{\chi}$. The flux limits are
calculated for the same muon minimum energy as is used in the
calculation ($E_{\mu} > 1$ GeV) considering the dependence
on the energy of the MACRO acceptance in the low energy region.
A correction is applied for each neutralino mass to translate
from the thresholds of Tab.~\ref{tab1}
to the 1 GeV threshold used in the calculation of upward-going muon fluxes.
The correction factors are higher for lower neutralino masses. Moreover,
they are higher for the Sun for which the threshold in Tab.~\ref{tab1} is
2 GeV than for the Earth for which it is 1.5 GeV. We estimate these factors 
to be 1$\%$ for the
Earth and 10$\%$ for the Sun for a neutralino mass of 60 GeV. 

The fluxes are calculated by varying
the model parameters (each model is represented by a dot in
Fig.~\ref{fig5} and in Fig.~\ref{fig6}) in experimentally allowed
ranges.  
Dots correspond to variations of parameters between
the following values:
$\tan\beta$ = 1.01, 2, 3, 10, 40, 50; $m_{A}$ = 65, 70, 75, 80, 85, 90, 
95, 100, 200, 300, 500 GeV, where $m_{A}$ is the mass of the pseudoscalar 
Higgs; $m_{0}$ = 150, 200, 300, 500 GeV, where $m_{0}$ is the
common soft mass of all the sfermions and squarks;
$A$ = -3, -1.5, 0, 1.5, 3 where $A$ is the common
value of the trilinear coupling in the superpotential for the bottom
and top quark ($A$ is set to zero for the first and second family);
$\left| \mu \right|$ and $M_{2}$ are varied between 10 and 500 GeV in steps
of 20 GeV.
The calculation also assumes a neutralino rms halo velocity of 270 km
s$^{-1}$, a halo escape velocity of 650 km s$^{-1}$, a velocity of
the Sun around the Galactic center of 232 km s$^{-1}$, a local dark
matter density of 0.5 GeV cm$^{-3}$, and a minimal value for rescaling
the neutralino relic abundance, $(\Omega h^{2})_{min}$, of 0.03
(explained below).  
Those fluxes lying above the experimental flux
limit curve are ruled out as possible SUSY models by this measurement
given the cosmological parameters chosen.
It should be considered that a variation of the astrophysical
parameters may lower the calculated fluxes by at most one order of
magnitude \cite{Bottino98}. Moreover, $\nu_{\mu} - \nu_{\tau}$
oscillations could lower neutrino fluxes from neutralino annihilations 
by about a factor of two for the oscillation parameters considered 
in this paper \cite{Fornengo}.

MACRO flux limits can be used to constrain also other neutralino
calculations, e.g. Bergstr\"om {\it et al.} \cite{Bergstrom97}. In this
calculation configurations with $\Omega_{\chi}h^{2} < 0.025$ are
not taken into account. On the other hand, Bottino {\it et al.}  
\cite{Bottino98},
when the neutralino cosmological abundance is too low to account for the
total dark matter in the halo, assume:

\begin{eqnarray}
\begin{array}{cl}
{\rm if}\; \Omega_{\chi} h^{2} > (\Omega h^{2})_{min} & \rho_{\chi} = 
\rho_{loc} \\
{\rm if}\; \Omega_{\chi} h^{2} < (\Omega h^{2})_{min} & \rho_{\chi} = 
\rho_{loc} \times  \Omega_{\chi} h^{2} / (\Omega h^{2})_{min} \, .
\end{array}
\end{eqnarray}

\noindent where $(\Omega h^{2})_{min}$ = 0.03 and  
$\rho_{loc} = 0.5$ $\rm GeV \, cm^{-3}$.
Both calculations indicate that indirect searches can have good prospects
particularly at high neutralino masses (see Fig.~\ref{fig5} and 
Fig.~\ref{fig6}). 

We have tested another approach to increase the sensitivity in the estimate
of flux limits for the Sun 
through the study of the angular distribution of the signal 
\cite{finestre}. The data, signal, and background angular distributions 
are fit in the region of interest for the signal, using a $\chi^{2}$ 
expression suited for Poisson-distributed data \cite{PDB}. In this expression
the signal is multiplied by a factor $K$. The normalization factor
$K$ of the signal and its error are found by minimizing $\chi^{2}$
with respect to $K$. The resulting flux limits are evaluated at the
90$\%$ c.l. as a function of the neutralino mass. The results of
this method, although still with limited statistics,   
are in agreement with the previous one in which
the flux limits are calculated in cones which collect 90$\%$ of the
signal, because our data in the direction of the Sun are in good
agreement with the prediction of the atmospheric neutrino background.
In principle, this method is eventually preferable because it avoids
an {\em a priori} choice of the search cone and hence
of the fraction of the signal lost outside of it
and it will improve with increasing statistics.

\begin{center}
\begin{figure}[htb]
\epsfig{file=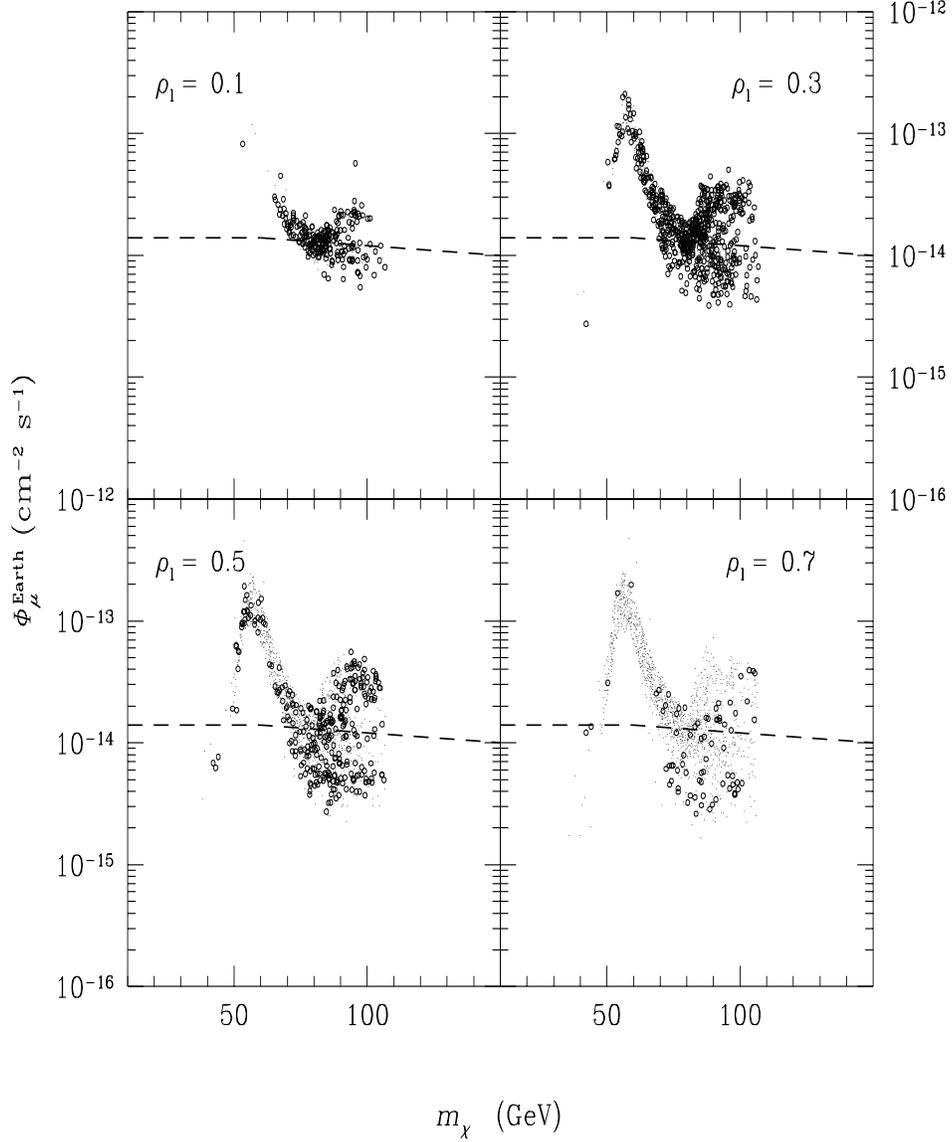,height=16.cm,width=13.cm}
\caption{The dashed lines are our experimental upgoing muon flux limits
(for $E_{\mu}^{\rm th} = 1$ GeV) at 90$\%$ c.l. as a function of 
$m_\chi$ for the Earth. The dots and the circles 
are SUSY neutralino models allowed at 90\%
c.l. from Bottino {\it et al.} assuming various local dark matter densities
\protect\cite{Bottinol}. The experimental flux limit and the theoretical 
fluxes shown are obtained assuming no neutrino oscillations. The limits 
computed assuming oscillations are  indistinguishable in this graphs; the
expected fluxes may decrease by about a factor of two. The dots 
represent models already excluded by cosmic anti-proton data
\protect\cite{Bess}.}
\label{fig7}
\end{figure}
\end{center}

Recently the DAMA/NaI experiment observes a possible a possible
annual modulation effect in a WIMP direct search at a 99.6$\%$
c.l. \cite{DAMA98}. This modulation has been interpreted in terms of
a relic neutralino which may make up the major part of dark matter in
the universe (see Ref.~\cite{Bottinol} and references therein). 
Fig.~\ref{fig7} shows the allowed SUSY models considered by 
Bottino {\it et al.} for various neutralino local density 
compared to the MACRO upward-going muon flux limits
from the Earth \cite{Bottinol}.
MACRO experimental upper limits from the Earth rule out many SUSY
configurations indicated by the DAMA/NaI experiment, even assuming the
atmospheric neutrino background oscillates with the parameters favored
by MACRO \cite{Letter} and Superkamiokande \cite{SK}. 
The MACRO data from the Sun, which is more sensitive 
to spin-dependent scattering,
have less overlap in sensitivity with direct searches.
We again notice that expected fluxes may decrease by about a factor of two
in the presence of neutrino oscillations.

\section{Conclusions}

A search for a WIMP signal is performed using the MACRO detector at
the Gran Sasso Laboratory to observe upward-going muons coming from the
products of WIMP annihilations from the Earth and Sun. The
time-of-flight technique is used to discriminate the upward-going
neutrino-induced events from the background of downward-going
atmospheric muons.

We look for an excess of neutrino events over the background of
atmospheric neutrinos in the direction of the Sun and Earth. No
signal of WIMP annihilation is observed, and we present upper limits
on the flux of upward-going muons from these bodies.

While we do not see a WIMP signal, we compare our flux limits to
predictions from various SUSY models and rule out those which are
inconsistent with our limits. We calculate the expected upward-going muon
fluxes and angular distributions using the SUSY neutralino
annihilation calculations of Bottino {\it et al.}
\cite{Bottino98}. The 90$\%$ c.l. upper limits for the flux of
upward-going muons in angular cones collecting 90$\%$ of the expected
signal from neutralino annihilation are given. Our data exclude
significant portions of the parameter space for neutralinos and our
limits for the Earth are the most stringent of all ``indirect''
experiments \cite{baksan96,Mori93}.
The flux limits for the Sun however tend to rule out a smaller 
portion of the SUSY parameter space at this time.
Finally, MACRO data also exclude some of the SUSY models
suggested by the annual modulation analysis performed by the DAMA/NaI 
experiment (especially at lower neutralino masses) \cite{DAMA98}. The MACRO
indirect search for WIMPs has reached the proper sensitivity to
explore configurations compatible with the DAMA/NaI data. 

\acknowledgments
We thank Prof. A. Bottino and Dott. N. Fornengo for valuable discussions
and for their contributions to this work. 
We gratefully acknowledge the staff of the 
{\it Laboratori Nazionali del Gran Sasso} 
and the invaluable assistance of the
technical staffs of all the participating Institutions. For generous 
financial contributions we thank the U.S. Department of Energy, the National
Science Foundation, and the Italian {\it Istituto Nazionale di Fisica 
Nucleare},
both for direct support and for FAI grants awarded to non-italian MACRO 
collaborators.

\begin{table}[htb]
\begin{tabular}{cccccccc}
 \multicolumn{1}{c}{ }&
 \multicolumn{4}{c}{EARTH}&
 \multicolumn{3}{c}{SUN} \\ \hline
 \multicolumn{1}{c}{Cone}  
&\multicolumn{1}{c}{Data}
&\multicolumn{1}{c}{Back-} 
&\multicolumn{1}{c}{Norm.}
&\multicolumn{1}{c}{Flux Limit} 
&\multicolumn{1}{c}{Data} 
&\multicolumn{1}{c}{Back-} 
&\multicolumn{1}{c}{Flux Limit} \\ 
 & & ground& factor & ($E_{\mu} > 1.5$ GeV) & & ground& ($E_{\mu} > 2$ GeV)\\
 & & events&        & (cm$^{-2}$ s$^{-1}$)  & & events& (cm$^{-2}$ s$^{-1}$)\\
\hline
$30^{\circ}$ &76& 119.2& 0.82& 2.24 $\times 10^{-14}$&56 &
51.8&5.84 $\times 10^{-14}$\\
$24^{\circ}$ &52& 75.5 & 0.79& 1.75 $\times 10^{-14}$&33 & 
33.1& 3.83 $\times 10^{-14}$ \\
$18^{\circ}$ &32& 42.2 & 0.77& 1.42 $\times 10^{-14}$&17 & 
18.8& 2.60 $\times 10^{-14}$\\
$15^{\circ}$ &24& 29.0 & 0.76& 1.20 $\times 10^{-14}$&11 & 
13.2& 2.09 $\times 10^{-14}$\\
$9^{\circ}$ &10& 10.5 & 0.75 & 7.48 $\times 10^{-15}$ & 3 &  
4.7& 1.34 $\times 10^{-14}$\\
$6^{\circ}$  &4 &  4.6 & 0.75& 5.30 $\times 10^{-15}$& 2 &   
2.1 & 1.37 $\times 10^{-14}$\\
$3^{\circ}$  &0 &  1.1 & 0.75& 3.79 $\times 10^{-15}$& 2 &  
0.5 &1.73 $\times 10^{-14}$\\
\end{tabular}
\caption{\label{tab1} Observed and background events and 90$\%$
c.l. muon flux limits for some of the 10 half-cones chosen pointing
from the Earth and the Sun. The background events are those expected
from atmospheric neutrinos. 
In the case of the Earth, the normalization factors are used
to normalize the expected background events 
by the ratio of observed to expected events outside each cone. 
Since the number of detected events (column 2) is less than the normalized 
expected events (column 3), we set conservative flux limits assuming that 
the number of measured events equals the number of expected ones 
\protect\cite{PDB}. 
The Earth results are for the no oscillation scenario. 
Earth limits considering neutrino oscillations agree to within 8$\%$. 
The average exposure for the Earth is 2620 m$^{2}$ yr and for the Sun
890 m$^{2}$ yr.}
\end{table}

\end{document}